\documentclass{aa}
\usepackage{natbib}
\usepackage{epsfig}
\usepackage{txfonts}

\def\f1{f_{\rm 1}}
\begin{document}
\title{Bulk composition of the transiting hot Neptune around GJ\,436}
\author{P. Figueira$^1$ \and F. Pont$^2$ \and  C. Mordasini$^3$ \and Y. Alibert$^{3,4}$ \and C. Georgy$^1$ \and W. Benz$^3$}
\offprints{pedro.figueira@obs.unige.ch}
\institute{$^1$ Observatoire de l'Universit\'{e} de Gen\`{e}ve, 1290 Sauverny, Switzerland\\
$^2$ University of Exeter, The Queens Drive, Exeter, Devon, England\\
$^3$ Physikalisches Institut, University of Bern, Sidlerstrasse 5, 3012 Bern, Switzerland\\
$^4$ Institut UTINAM, CNRS-UMR 6213, Observatoire de Besan\c{c}on, BP 1615,
25010 Besan\c{c}on Cedex, France}
\date{Received date / accepted date}

   \authorrunning{P. Figueira et al.}
   \titlerunning{Bulk composition of the transiting hot Neptune around GJ\,436}
\abstract{The hot Neptune orbiting around GJ\,436 is a unique example of an intermediate mass planet. Its close-in orbit suggests that the planet has undergone migration and its study is fundamental to understanding planet formation and evolution. As it transits its parent star, it is the only Neptune-mass extrasolar planet of known mass and radius, being slightly larger and more massive than Neptune (M=22.6 M$_\oplus$, R=4.19R$_\oplus$). In this regime, several bulk compositions are possible: from an Earth-like core with a thick hydrogen envelope to a water-rich planet with a thin hydrogen envelope comprising a Neptune-like structure. We combine planet-structure modelling with an advanced planet-formation model to assess the likelihood of the different possible bulk compositions of GJ\,436\,b. We find that both an envelope-free water planet (``Ocean planet'') as well as a diminute version of a gaseous giant planet are excluded. Consisting of a rocky  core with a thick hydrogen/helium envelope, a "dry" composition produces not only too small a radius but is also a very unlikely outcome of planet formation around such a low-mass star. We conclude that GJ\,436\,b is probably of much higher rock content than Neptune (more than 45\% in mass), with a small H-He envelope (10 - 20\% in mass). This is the expected outcome of the gathering of materials during the migration process in the inner disk, creating a population of which the hot Neptune is representative.

}

\maketitle

\section{Introduction}

%
%

The 22-Earth-mass planet orbiting around the late M dwarf GJ\,436 (Butler et al. 2004, 
Maness et al. 2007) transits its parent star (Gillon et al. 2007a). Photometric 
monitoring of the transits from the ground and with the Spitzer Space Telescope inferred a
planetary radius of 4.19$^{+0.21}_{-0.16} R_\oplus$ (Gillon et al. 2007b). It is the first intermediate-mass planet for which we have a radius measurement and is found at an orbital distance that implies that it has been affected by migration. It provides us with a unique test-bed to study the formation, evolution and migration of planets.  For the first time we are able to analyze the different characteristics of an exoplanet of this class.

In the current paradigm of planet formation and composition -- still heavily based on 
the single case of the Solar System -- planets consist of four main categories of compounds: 
H/He, ices, silicates and iron/nickel\footnote{Kuchner et al. (2003) also consider the case 
of ``carbon planets'', planets with a carbon-to-oxygen ratio higher than unity. Until the 
observations indicate otherwise, we can assume that the C/O ratio around GJ\,436 is 
normal, since the object is an otherwise unremarkable M-dwarf in the solar 
neighbourhood.}. Planets consisted almost entirely of only one of these four materials have very different radii due to the wide range in densities. Since different materials are mixed in the same planet, a certain amout of degeneracy is unavoidable due to the fact that a mass-radius pair can be formed by different arrangements of elements. Nevertheless, the observed mass and radius of a planet can provide valuable constraints on its bulk composition. 

The mass-radius combination for GJ\,436\,b suggests that the planet consists primarily of heavy 
elements (iron/rocks/ices), surrounded by a lighter envelope -- just like Uranus and 
Neptune.  Given its tight orbit, it is unlikely to have formed close to its present location and is therefore  supposed to have undergone type I migration  within the protoplanetary disc. Its proximity to its parent star implies that it may also have been affected by evaporation.  
We explore all the possible composition scenarios and address three key questions requiring a finer analysis than presented so far:

\begin{itemize} 
\item{how likely is the planet to be a ``dry'' neptune, i.e. to have a core of refractory elements that 
accreted a H/He envelope inside the snow line, without accumulating ices in significant amounts?}
\item{could it be an ``ocean planet'', i.e. a water-dominated planet, whose H/He envelope has been lost by evaporation?}
\item{could its composition  be significantly different from that of Neptune due to its peculiar formation process?}
\end{itemize}


To attempt to answer these questions, we combine simple but realistic planetary structure models with the formation models of Alibert et al. (2005) and Mordasini et al. (2008a, 2008b). In Sect. 2 we present our structure model and in Sect. 3 we review the characteristics of the formation model employed. Section 4 describes the results obtained using the two models, first separately and then by using the models together. We discuss our results in Sect. 5 and present our conclusions in Sect. 6.

\section{Structure model}

The discovery by radial velocity surveys of close-in planets in the 4-20 Earth-mass range (e.g. Lovis et al. 2004, Udry et al. 2007,  Mayor et al. 2008) has motivated the development of several sets of 
structure models of these planets (Valencia et al. 2007; Fortney et al. 2007; Seager et 
al. 2007). As emphasized by these authors, the mass-radius relation is a robust 
prediction of the structure models, far less sensitive to details and model-input 
uncertainties than, for instance, predictions about the thermal history or atmospheric 
spectrum of exoplanets. Sasselov et al. (2007) discussed the effects of the different 
sources of uncertainties on the radius predictions, and found that the most significant effect -- 
the equation of state (see review by Guillot 2005) -- produced an error of 2\% in the radius. This is smaller than the observational uncertainties in most cases (the radius uncertainty is 4\% for 
GJ\,436\,b). Therefore, a relatively simple model is sufficient to interpret the position of 
GJ\,436\,b in the mass-radius diagram in terms of composition. 
The relative insensitivity of radius predictions to the details of the structure models is due to the large spread in density between the different raw materials (spanning almost two orders of magnitudes from hydrogen to iron), and the cubic relation between the mean density and the radius. An error of 10 \% in the equation of state in a certain regime will produce only a 3.3\% difference in the radius of the layer concerned.

The authors mentioned above have concentrated generally on the $M<10 M_\oplus$ regime, 
because these masses are too small for the planet to accrete a significant gaseous envelope. 
Therefore, the correspondence between planetary radius and composition is more direct. 
Nevertheless, even if an unequivocal determination of the composition is impossible 
for GJ\,436\,b, educated guesses can be made from the same type of structure models.

We developed a structure model for Neptune-type planets close to their parent star 
(``hot neptunes''), and present an application to the case of GJ\,436\,b. Our approach closely follows that of Fortney et al. (2007).
 We use four layers in our model: iron/nickel, silicates, ices, and hydrogen/helium. We use the equations of states 
from Fortney et al. (2007) for iron and silicates at zero temperature. As suggested by 
these authors, we neglect the small thermal expansion factor for iron and silicates. We 
use the ``hot water'' equation of state from Fortney et al. (2007) for the ices layer, including a 
first-order account of the thermal effect for a reasonable temperature profile. 
For the hydrogen/helium mixture, we use the equation of state of Saumon et al. (1995) of a mixture of 75\% hydrogen and 25\% helium, with a temperature profile taken along 
an adiabat. We define the outer limit of our structure models as being at a pressure of 1 kbar. The thickness of the outer layer, from 1\,kbar outwards, is estimated with a different procedure (see below).

Depending on the exact temperature history of the formation of the initial planetary 
core, a certain amount of ammonia and methane can be mixed with the water. 
To estimate the importance of a potential enrichment in these non-water ices, we 
also consider a suite of models with the mixed water+ammonia+methane, using the density ratio of pure water ice to mixed ice from the Neptune/Uranus models of Hubbard et al. (1995).
Since it is not known, even in Neptune and Uranus, whether the refractory elements 
are mixed with the water layer or separated in a central core, it is not useful in this 
mass regime to attempt to refine the equations of state beyond a certain point.
Moreover, the pressure at the centre of a 22-Earth-mass planet (10-100 Mbar) exceeds
the point reached by high-pressure experiments in laboratories considerably for the relevant 
compounds, so that a certain degree of uncertainty is unavoidable. The key point is that the uncertainties in the model radius remain smaller than the observational uncertainties in the radius of the planet.

The thickness of the outer layer, from 1 kbar to the pressure at which the optical depth becomes small, has a stronger dependence on the temperature profile. We define the ``transit radius'' as being that at which the pressure is 0.1\,bar. There is some level of uncertainty in this value, and it is also wavelength-dependent. We compute the thickness of the outer layer by extending the measured pressure-density profile and scale height of the atmosphere and outer layers of Neptune (Podolak 1976) to the hotter temperature of GJ\,436\,b.  Between 0.1 and 1 bar (``atmosphere''), we adjust the scale height according to the ideal gas law H $\sim$ T/$\mu$g.
We use the equilibrium temperature of GJ\,436\,b ($\sim$600 K) as an indication of the temperature in its isothermal layer.
Between 1 bar and 1 kbar (``outer layer''), we follow the analytical treatment of Arras \& Bildstein (2006) to modify the temperature profile for a higher temperature in the isothermal layer.

This modelling is approximate and does not provide realistic predictions of the atmosphere' thickness of the planet. The typical thickness of the 1kbar to 0.1 bar zone in our models with a H/He envelope is around 1$\times$10$^{3}$\,km, or 4\% in terms of radius. Therefore, even if these estimates are uncertain, they will affect the total radius only at the percent level, which is acceptable for our purposes. 

For models without an H/He envelope, we proceed as follows: for water (steam atmosphere), we use the same approach as for H/He, scaling according to an ideal gas law for the difference in mean molecular weight. In an alternate model, we use a temperature of 1$\times$10$^{3}$\,K for the atmosphere, to account for the possibility of a strong greenhouse effect elevating the temperature of the isothermal layer several hundred degrees above the equilibrium temperature. If the outermost layer is silicates or iron, then we fix the density of the outer layer to the zero-pressure density of the solid state of the material (in that case the thickness of the outer layer has a negligible effect on the total radius).
 
After the equations of state have been chosen, the mass and density profiles can be calculated.
The effect of the temperature profile in our models is integrated into the equations of 
state, and we do not solve the equation of energy transfer. We are left with two basic 
structure equations: the continuity condition

\begin{equation}
\hspace{5 mm}\frac{\partial r}{\partial M_r} = \frac{1}{4\pi r^2 \rho}\label{cont}
\end{equation}

\noindent and the hydrostatic equilibrium

\begin{equation}
\hspace{5 mm}\frac{\partial P}{\partial M_r} = -\frac{G M_r}{4\pi r^4}\label{equ}
\end{equation}

\noindent where $M_r$ is the included mass at the level $r$ and $P$, $\rho$ the pressure and density at that level.

Equations \ref{cont} and \ref{equ}, in addition to the equation of state, form a full system in which all the variables are determined at each level. For the integration, we use a standard shooting method (eg. Press et al. 1992), beginning at the center and going towards the surface. We start with an arbitrary central pressure, and compute the surface value. Depending on the result, we modify successively the central value, until the surface value reaches the required value. The mass fraction of the different layers is an input parameter, which enables the composition parameter space to be completely explored.

Our structure model was tested with Solar System objects. With accepted 
composition mixtures, we were able to reproduce the radius of Mars, Earth, the Moon, 
Titan, Uranus and Neptune to a closer margin that a few percents, usually about 1\%. This does 
not prove that the composition mixtures adopted were correct, since they were also based 
on models developed under the same assumptions. It demonstrates, however, that no critical property cannot be reproduced by our simplified approach. We compared our model results with the equivalent of Fortney et al. (2007) and Seager et al. (2007) for the mass range that we explored. The comparison with the former shows that the output radii always agree to within 4\% and the differences most of the time are around 1\%. When we apply the polytropic state equations of Seager et al. (2007), we recover their radii to within 1\%. Once again, the numerical approach of the model is validated.

The structure models employed here are far too simple to take into account effects such as the radioactive decay of elements or excitation pumping. We decided against trying to include these particular effects because it would have increased the degeneracy of the solutions, due to the insufficient amount of currentl available data.

The inputs to our structure model are the mass fraction of iron, silicates, ices, and 
H-He, the total mass, and the surface pressure. The outputs are the mass, pressure and density profiles, and the total radius.
We ran the model for a variety of composition mixtures, sampling the parameter 
space around the observed mass-radius position of GJ\,436\,b. 

\section{Formation Model}

As stated in the previous section, the measured mass and radius of GJ\,436\,b provide constraints
on its internal structure. On the other hand, the formation process itself
establishes the final composition of the planet. The bulk internal composition must be consistent not only with the measured radius and mass of GJ\,436\,b but also with a plausible formation scenario.

We use here the latest version of the extended core-accretion models including
protoplanetary disk structure and evolution, as well as migration of the forming planet.
Details of the formation model itself can be found in Alibert et al. (2005), whereas
information about the initial conditions that we consider here are described in Mordasini et al. (2008a).
In short, the calculation of the formation includes, in a consistent way, three effects:
1) the accretion of solids and gas, 2) the migration of the planet, and 3) the progressive
disparition of the protoplanetary disk, due to viscosity and photoevaporation.
The accretion of solids and gas is calculated as in standard core-accretion models
(e.g. Pollack et al. 1996). The migration of the planet occurs within two regimes: low mass
planets migrate within type I, whereas higher mass planets migrate within type II. The switch
from type~I to type~II occurs when the planet's hill radius exceeds the local disk thickness.
However, for the low-mass planets considered here, type II migration is not important.
The type I migration rate is calculated using the analytical work of Tanaka et al. (2002),
reduced by a constant factor $\f1$, which stems from different effects that can in fact
reduce the migration velocity compared with the value derived for the ideal case considered
in Tanaka et al. (2002). These effects include in particular magnetic induced turbluence
(Nelson \& Papaloizou 2004), or non-isothermal effects (Menou \& Goodman 2004; Paardekooper
\& Mellema 2006). To reproduce the bulk internal composition and atmospheric composition
of Jupiter and Saturn, as well as the statistical properties of extrasolar planets, values of
 $\f1$ between 0.001 and 0.1 are required, as established by Alibert et al. (2005). The contemporary work of Mordasini et al. (2008b) pushed this approach further, and studied the effect of not only $\f1$  but also of the dust-to-gas ratio and the disk viscosity parameter, among others, on the final population properties. In the end, a coherent picture was attained in which the observable properties of extrasolar planet population, such as mass, semi-major axis and ``metallicity effect" (increase of detection probability with metallicity) were reproduced. The models presented here stand on this parametrization and assumptions, and in particular $\f1$ takes the value of 0.001.

For the case of the GJ\,436 system, we considered formation models around a
0.5 M$_\odot$ star with a solar or half-solar ([Fe/H]=$-0.3$) metallicity, reflecting
the uncertainties in observations (Maness et al. 2006, Bonfils et al. 2005). We have calculated
tens of thousands of formation models, each one assuming a different set of initial conditions,
namely the mass and lifetime of the disk, and the initial location of the embryo that will
(eventually) lead to a planet. The statistical distribution of the disk lifetime is taken
from Haisch et al. (2001), whereas the distribution of disk mass is scaled down from
observations around solar mass stars, using the relation $M_{disk} \propto M_{star}^\beta$ 
with $\beta \sim 1.2$ (Beckwith \& Sargent 1996). This latter relation allows us to calculate accretion rates in computed disks, which evolve with the square of the star mass, as indicated by observations of protoplanetary discs (Muzerolle et al. 2003; Natta et al. 2004).

Although a scaling of the mass of the disk with the host star mass is
theoretically expected from the gravitational link between these quantities,
observations do not support such a clear picture. For example, Andrews \&
Williams (2005) demonstrated that from submilimeter continuum slopes in
Taurus Auriga one cannot infer a correlation between $M_{star}$ and $M_{disk}$
for the interval of $M_{star}$ of interest here. As a limiting case, we therefore considered the scenario in which the disk mass does not scale with $M_{star}$ between FGK and M stars i.e. $\beta$\,=\,0.

\subsection{Evaporation}

\label{evap}

Our formation models terminate when the disc evaporates, and do not follow the subsequent evolution of the planet. In particular, subsequent evaporation of part of the planet atmosphere could modify its radius and bulk composition. Mass loss through atmospheric evaporation is thought to affect close-in planets, and could even strip a planet of most of its mass (e.g. Baraffe et al. 2006). 

However, comparison with other known close-in planets suggests that mass loss has not been a strong factor in the evolution of GJ\,436\,b. Despite its proximity to the parent star, the planet receives far less incident flux than hot Jupiters such as HD\,209458\,b, due to the low luminosity of the star. GJ\,436\,b occupies a position in the comparison of the potential energy at the top of its atmosphere with the incident stellar flux (see Lecavelier 2007), which implies that a significant part of its mass has not escaped through atmospheric evaporation.

\section{Results}

\subsection{Structure}

\begin{table*}
\centering
\begin{tabular}{lcccccc} \hline
 \ \ &  Jovian planet  & Ocean planet & Hot Neptune 1  & Hot Neptune 2 & Dry Neptune  \\ \hline
H/He   & 0.40   & 0.00  & 0.10  & 0.10  & 0.20  \\
Ices  & 0.30  & 0.80  & 0.60 & 0.40 & 0.00  \\
Rock  & 0.20   & 0.10  & 0.20  & 0.33  &  0.53 \\
Iron  & 0.10   & 0.10 & 0.10  & 0.17  & 0.27 \\ \hline
Mass [M$_\oplus$]& 22.6  &  24.5  & 22.6  & 22.6  & 22.6   \\
Radius [km]&  36596 &20833  &  27355  &  25929  & 27096  \\ \hline
\end{tabular}
\caption{Bulk composition and computed radius for some key models. Note that the GJ\,436\,b measured mass and radius are 22.6$^{+1.9}_{-1.9}$M$_\oplus$ and  4.19$^{+0.21}_{-0.16} R_\oplus$ ($\sim$26500$^{+1340}_{-1020}$ km) according to Gillon et al. (2007b).}
\end{table*}

Several structure models can be excluded at the 2-$\sigma$ level. The observed radius of GJ\,436\,b is not consistent with being a planet with more than 30\% in mass of light gases (small Jovian planet) or 90\% of rock and iron (a ``super Earth''). 

In the absence of H/He, with a purely ``Ocean planet'' model, neither,we were also unable to reproduce the observed radius. Models without a H/He envelope predict radii that are smaller than the observed radius by at least 20\%. Even when replacing water ice entirely with methane ice, adding a hot steam atmosphere at 1$\times$10$^{3}$K, and using the upper end of the observed mass range, the calculated radius for hydrogen-free planets with the mass of GJ\,436\,b does not reach the 1-$\sigma$ lower error bar of the observed radius. The minimum amount of H/He required is 5\% in mass, with a purely water core. 

Up to this point, our conclusions correspond to those already presented in Gillon et al. (2007a).

At the other extreme, a ``dry'' composition with a rocky core of 18.3 $M_\oplus$, and 16\% H/He in mass ($\sim50$ \% in radius) is also consistent with the observations. A higher H/He proportion produces a larger radius than observed. Changing the silicate to iron ratio from 2:1 to 1:1 modifies these results only at the 1.5 \% level, corresponding to a $\sim$ 350 km shrinkage in radius for the most likely bulk compositions.
 
Finally, many intermediate models, with a certain amount of ices compensated by lower amounts of light gases to retain the radius at a constant value, are compatible with the observed mass and radius. 

Table 1 provides the composition, radius, and mass for five models with four fundamentally different bulk 
compositions: the ' ``Jovian planet'', the ``ocean planet'', ``hot Neptune'', and ``dry Neptune'' options. Only the last three provide results that are compatible with the mass and radius observations. For the ``hot neptune'' variety, two different compositions are given, one with a majority of ices, the other with a majority of heavy elements.

These results are in greement with those of  Adams et al. (2007), as expected since our structure models are very similar.
 
\subsection{Formation}

\begin{figure}
\resizebox{8.8cm}{!}{\includegraphics{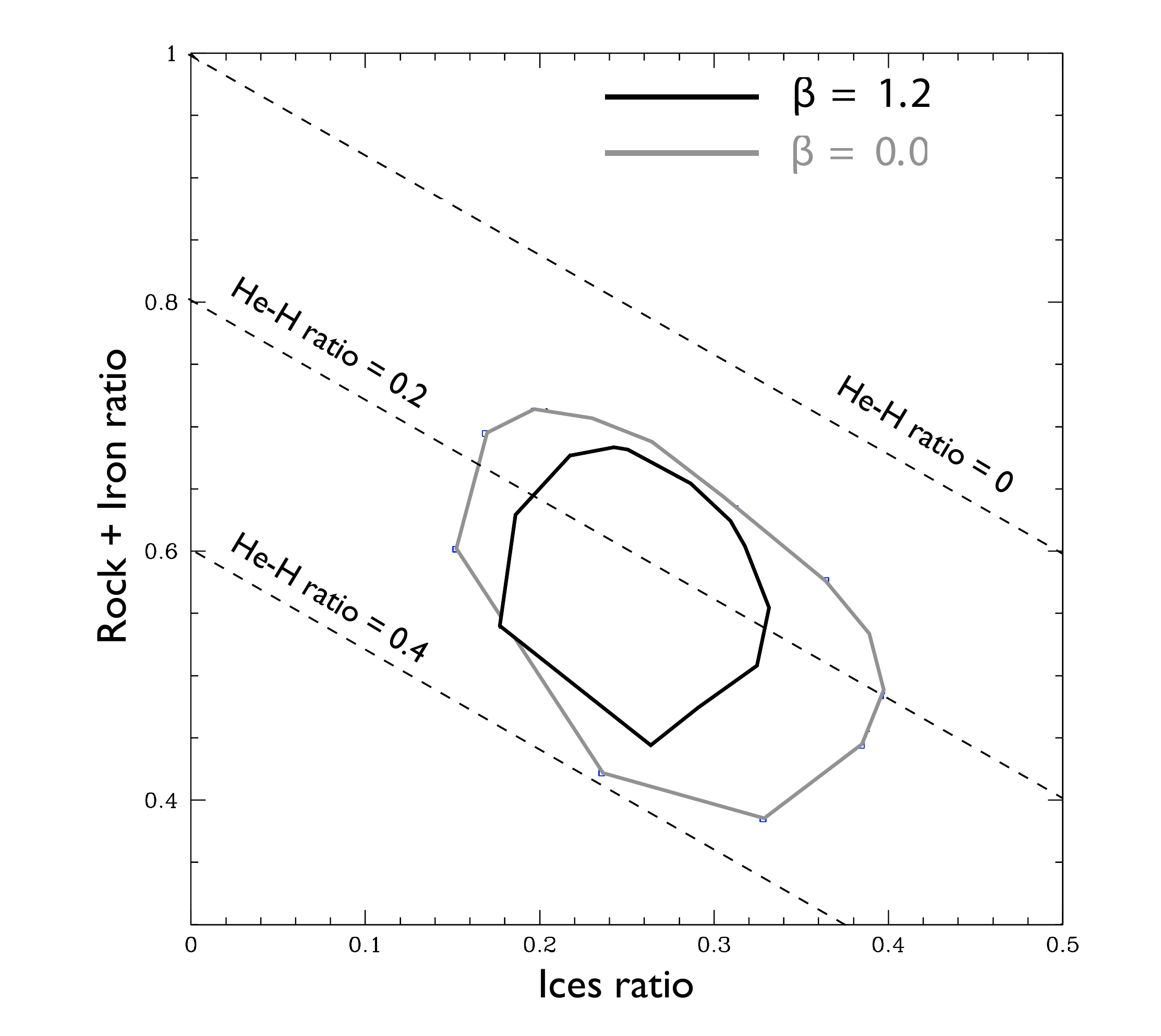}}
\caption{Internal composition of intermediate mass planets (22.6 $\pm$ 5 $\mathrm{M}_\oplus$) at low distance from the central star (below 0.1 AU). The two
curves delimit the envelope of the corresponding planets in two cases: $\beta = 1.2 $ in black, $\beta = 0.0$ in blue.
For each cases, two metallicities are considered, namely solar and sub-solar metallicity ($[Fe/H]$\,=\,-0.3). The regions occupied in the two cases are however very similar, and are not
distinguished in the figure for clarity.  The lines in dash
are lines of constant gas mass fraction (value indicated on the figure).}\label{compo}
\end{figure}

We calculated different sets of formation models, differing in value of $\beta$ and the metallicity of the system (solar or half solar). In each case,
we selected planets whose final location was $<$ 0.1 AU (close-in planets), and total mass in the range 17.6--27.6 $M_\oplus$. For these planets, the internal composition (fraction of ices,
refractory elements and H/He) was given by the formation model. The formation models do not consider the evolution of planets after their formation, and in particular possible further evaporation (see Sect. \ref{evap}). The effects of the presence of multiple embryos and giant impacts are also not considered by the model (for a discussion of the importance of these effects the reader is referred to Ikoma et al. 2006 and Thommas et al. 2008, respectively). 
We emphasize that in our models, the formation parameters are established to provide a good match for the extrasolar planets' properties around solar mass stars (see Mordasini et al. 2008b).
Figure \ref{compo} illustrates the composition of planets satisfying the mass and orbital distance criteria, for different values of the $\beta$ parameter, which was left as a free parameter for the reasons mentioned in Sect. 3.

We note that the 17.6\,--\,27.6 $M_\oplus$ mass range corresponds to the most common outcome of the models for a 0.5 $M_\odot$ star mass. Therefore, the formation of GJ\,436\,b can be explained by the models, which predict that it has a high probability of ocurring, i.e. as the element of a typical population and not as a single event case.
 
\subsection{Formation and structure}

We now apply our structure model to calculate a radius for each of the planets produced by the formation model. Since the formation model does not distinguish between rocks and iron components, we use a 2:1 rock-to-iron ratio. As stated earlier, using other realistic ratios does not significantly affect the radius predictions.

\begin{figure*}
\centering
\resizebox{180mm}{!}{\includegraphics{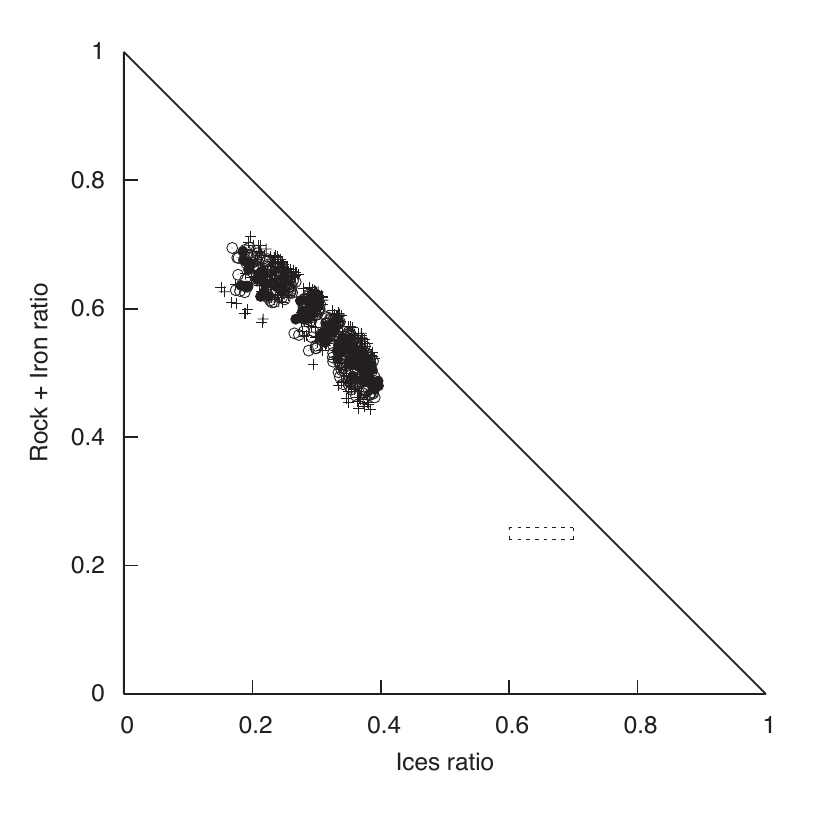}\includegraphics{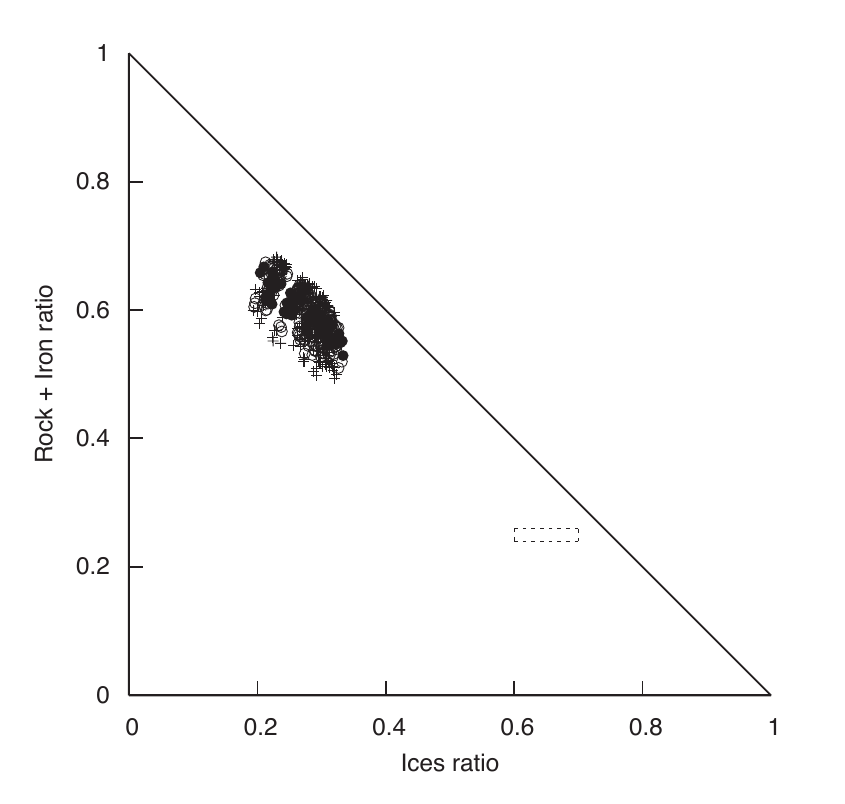}}
\caption{Internal composition of intermediate-mass planets (18.6\,--\,27.6 $\mathrm{M}_\oplus$) at low distance from the central star (below 0.1 AU) in the formation models for solar metallicity. The composition of synthetic planets is expressed in terms of rock + iron and ices ratios. The left panel corresponds to the planets created using $[Fe/H]$\,=\,0.0 and the right $[Fe/H]$\,=\,-0.3; in both simulations we assumed that $\beta$\,=\,0, since the results do not depend strongly on this parameter (see Table \ref{popul} for details).  The dotted box corresponds to the composition of Neptune if we assume a protosolar ice/rock ratio (as considered in Guillot, 1999). The symbol types indicate the level of agreement of the computed radii and mass with those measured for GJ\,436\,b. Filled circles represent simulated planets within the estimated modelling errors from the observations and open circles within twice these errors. Crosses represent all other models located within 3 times the estimated errors for mass and radius. }\label{compo2}
\end{figure*}

To evaluate the likelihood that a given synthetic planet corresponds to GJ\,436\,b, we compare the model outputs with the measurements of planet mass and radius (Gillon et al. 2007b) simultaneously.  This allows us to determine the parameter space that can reproduce both measurable quantities. However, when selecting the synthetic planets that correspond to the observations, one has to take into account not only the observational errors but also the uncertainties  present in the modelling itself. We already estimated our radius modelling errors to be inferior to 4\% (about 0.17R$_\oplus$), which we can consider to be a conservative upper limit. In what concerns mass uncertainties we have to note that  the formation model does not include evaporation or subsequent dynamical evolution of the system, including possible mergers. In the context of mass uncertainties, we recall that the formation model does not include evaporation or subsequent dynamical evolution of the system, including possible mergers. Besides that, the precise final mass depends on the disk evaporation timescale which therefore determines an end to the evolution of the planets. To take into account these uncertainties we introduce a conservative $\sim 10$\% margin, or 2$M_\oplus$, in the comparison between models and observational determination of planet mass. We selected all of our synthethic planets within 0.17R$_\oplus$ and 2$M_\oplus$ from the measured radius and mass, as candidates to represent GJ\,436\,b internal composition. A second group within twice the above margins is also selected and analysed. We obtain aproximately the same final composition, which demonstrates that the results do not depend on a precise estimation of the modelling uncertainties, as we can see in Table \ref{popul}. The resulting planetary population is shown in Fig. \ref{compo2}, with the different groups represented. 
 
\begin{table*}
\centering
\begin{tabular}{lccccc} \hline
 \ \ Model parameters & Model unc. & Rock+Iron ratio range & Ice ratio range & H-He ratio range \\ \hline
$[Fe/H]$ = 0.0;  $\beta$ = 0.0 & 1-$\sigma$ & 0.47 - 0.69  & 0.18 - 0.40  & 0.10 - 0.18  \\ 
\hspace{12 mm} --- & 2-$\sigma$ & 0.46 - 0.70 & 0.17 - 0.40 & 0.08 - 0.20  \\
$[Fe/H]$ = 0.0;  $\beta$ = 1.2 & 1-$\sigma$ & 0.48 - 0.70  & 0.17 - 0.39  & 0.10 - 0.16 \\ 
\hspace{12 mm} --- & 2-$\sigma$ & 0.47 - 0.70  & 0.17 - 0.39  & 0.09 - 0.19 \\ 
$[Fe/H]$ = -0.3;  $\beta$ = 0.0 & 1-$\sigma$ & 0.53 - 0.67  & 0.20 - 0.33  & 0.11 - 0.17  \\ 
\hspace{12 mm} --- & 2-$\sigma$ & 0.51 - 0.67 & 0.20 - 0.33 & 0.09 - 0.20  \\
$[Fe/H]$ = -0.3;  $\beta$ = 1.2 & 1-$\sigma$ & 0.53 - 0.66  & 0.21 - 0.33  & 0.10 - 0.16 \\ 
\hspace{12 mm} --- & 2-$\sigma$ & 0.52 - 0.68  & 0.19 - 0.33  & 0.09 - 0.20 \\ \hline
\end{tabular}
\caption{Bulk composition of synthethic planets for the measured mass and radius of  GJ\,436\,b, for different $\beta$ and [Fe/H]. The planets composition within an interval centred on its measured value and of width given by a factor one or two the modelling uncertainties are shown.}
\label{popul}
\end{table*}

The synthetic population is remarkably homogeneous in composition, as the clumping of data points in Fig. \ref{compo2} testifies. As reported in Table \ref{popul}, the assumption of different $\beta$ scaling factors produces minimal population changes. The robustness of the results to the different physical parameters variations allows us to draw some conclusions about the bulk composition of GJ\,436\,b. 

First, no model reproduces the observed orbital distance and mass without including a significant amount of water in the composition. Thus, each and every embryo that evolved into the planets of the selected population was formed beyond the ice-line. Pure rock models either migrate too rapidly to accumulate sufficient mass before reaching the close neighborhood of the star, or migrate too slowly and reach runaway gas accretion creating a gas giant. Retaining a ``dry neptune'' in the 17-28 $M_\oplus$ mass range requires a fine-tuning that was never achieved in our simulations. Planets without H-He envelopes (``ocean planets'') also never occur in the models, because a significant amount of gas is always captured from the disc during the migration, even without runaway accretion. The composition of the model planets on close-in orbits and in the correct mass and radius range from 17\% to 40\% of their mass in the form of water, 10-20\% in the hydrogen-helium envelope, and 45\%-70\% in the rock+iron core. 

\section{Discussion}

The precise radius of GJ436\,b measured by Gillon et al. (2007b) and Deming et al. (2007) suggested a Neptune-like bulk composition for this planet. This paradigm for the planet is not, however, confirmed by our work. 
Using our structure models we have demonstrated that, in the current planet composition scenario, an envelope of light gases amounting to 10-20\% in mass should indeed be present. An envelope-free planet would be substantially smaller, even assuming a high proportion of methane and ammonia combined with the water ice and a hot greenhouse atmosphere. With this quantity of light gases, the radius of GJ\,436\,b is consistent with all ratios of water to refractory elements, from a purely water core to a purely dry core. 

The combination of the structure and formation models yield water contents of between 17\% and 40\% in mass for planets such as  GJ\,436\,b. Rock and iron should therefore constitute more than half the bulk mass of the planet, unlike Neptune and Uranus which contain a higher fraction of ices: 25\% rock+iron, 60-70\% ices and 5-15\% H-He (Guillot 1999; Podolak et al. 2000). This is because, although the planet begins its formation beyond the snow line, it accumulates a substantial fraction of its mass in regions of the disc closer to the star, where the fraction of refractory elements is higher when compared with the outer regions of the disk. As a result, we measure a composition for GJ\,436\,b that differs significantly from that of Neptune: the rock+iron ratio is probably between 45\% and 70\%, the ices ratio between 17\%, and 40\% and the H-He ratio between 10\% and 20\%. 

The migration of a Neptune-mass planet and consequent accretion are constrained by the parameter $\f1$, and different values may correspond to a different final structure. Even though this variable was fixed by Mordasini et al. (2008b), we decided to test its impact on the final structure. We calculated additional models using $\f1$=0.01 and 0.1 (using solar metalicity and $\beta = 1.2$) and compared their results with those obtained previously. While the fraction of H-He remained approximately constant, the amount of rock+iron ranged from 40\% to 80\% for $\f1$=0.01 and from 40\% to 95\% for $\f1$=0.1. This was expected since the protoplanets can be drawn to the inner limit of the numerical disk far more rapidly, creating a wider range of rocky contents at the expense of a reduced ice content, which does not have enough time to develop. It is therefore clear that the conclusions reached about the richness of the rock and iron content in GJ\,436\,b are not the result of fine-tunning the $\f1$ parameter but a typical outcome of a migration process within the inner part of the disc.

It is important to note that we have assumed that separated layers of different materials in our composition exist, which may not be the case. Using separated layers ususally produces rock-poor solutions when compared with models that allow a progressive mixing. Once again, this reinforces our belief in the high rocky content of the planet.

\section{Conclusion}

We have combined formation and structure models to narrow the composition parameter space that corresponds to the mass and radius measurements of a transiting planet. The strength of this method is its ability to reject composition scenarios providing predictions that remain plausible for structure models alone, independently of their sophistication. This problem is caused by the span in densities of different compositions and the cubic dependence of the radii on the density. By using a planetary formation model to provide realistic outputs from the formation process, this degeneracy can be significantly reduced. 

Our study has demonstrated that GJ\,436\,b can be described within the current paradigm of core accretion planet formation, including type I migration and the dependence of planet formation on the properties of the parent star.  We propose that this planet is not a Neptune analogue but instead a planet of higher rock and iron content (above 45\%), due to the accumulation of these compounds during its migration within the inner disk. For most of our study we used the nominal $\f1$ parameter value of 0.001, as established by Mordasini et al. (2008b). Since this parameter is poorly constrained theoretically, we tested that our results are not sensitive to its variation. As we experimented with other plausible values and the migration rate increased, the iron and rock content of GJ\,436\,b were also found to increase, reaching 95\% in mass for some planets, which corresponds to a ``dry Neptune" configuration. We are therefore convinced that the rocky nature of the hot-Neptune around GJ\,436 is not an artifact of our model or a consequence of our parametrization but a natural outcome of the formation process depicted.

Further detections of low-mass transiting planets are of course required to establish robust statistical conclusions. With the results of several ongoing or planned programs, such as the space missions {\it CoRoT} and {\it Kepler}, and ground-based radial-velocity planet surveys, these kind of studies can be performed. By analysing a low-mass planet population, we will be able to constrain the floating parameters of the formation models and thus expand our understanding of the underlying physics.

\begin{acknowledgements}
P.F. thanks Stephane Udry, Francesco Pepe and Michel Mayor for enlightening discussions and comments on a draft version of the paper. Support from the Funda\c{c}\~{a}o para Ci\^{e}ncia e a Tecnologia (Portugal) to P. F. in the form of a scholarship (reference SFRH/BD/21502/2005) is gratefully acknowledged. C.M. gratefully acknowledges the support by the Swiss National Science Foundation. Computations were made in part on the ISIS and ISIS2 clusters at the University of Bern.
\end{acknowledgements}

\end{document}